\documentclass[12pt]{article}

\usepackage{amsmath}
\usepackage{amssymb}
\usepackage{epsfig}

\newcommand{\capdef}{}
\newcommand{\mycaption}[2][\capdef]{\renewcommand{\capdef}{#2}%
        \caption[#1]{{\itshape #2}}} 
\makeatletter
\renewcommand{\fnum@table}{\textbf{\tablename~\thetable}}
\renewcommand{\fnum@figure}{\textbf{\figurename~\thefigure}}
\makeatother

\newcounter{myenumi}

\renewcommand{\themyenumi}{\roman{myenumi}}
{\end{list}}

\newlength{\myem}
\newcounter{mysubequation}[equation]

\makeatletter
\renewcommand{\section}{\@startsection{section}{1}{0em}{-\baselineskip}%
{\baselineskip}{\normalfont\large\bfseries}}
\renewcommand{\subsection}%
{\@startsection{subsection}{2}{0em}{-0.7\baselineskip}%
{0.7\baselineskip}{\normalfont\bfseries}}
\makeatother

\newcommand{\bea}{\begin{eqnarray*}}
\newcommand{\eea}{\end{eqnarray*}}

\newcommand{\eV}{\,\mathrm{eV}}

\newcommand{\reub}{{\bar{\nu}_e\rightarrow\bar{\nu}_\mu}}
\newcommand{\ruu}{{\nu_\mu\rightarrow\nu_\mu}}

\newcommand{\NKT}{{N_{\text{kT}}}}
\newcommand{\dm}[1]{{\Delta m^2_{#1}}}

\begin{document}


\begin{titlepage}

\renewcommand{\thefootnote}{\alph{footnote}}

\ \vspace*{-3.cm}
\begin{flushright}
  {\hfill TUM--HEP--375/00}\\
  {MPI--PhT/2000-20}\\
  {\ }
\end{flushright}

\vspace*{0.5cm}

\renewcommand{\thefootnote}{\fnsymbol{footnote}}
\setcounter{footnote}{-1}

{\begin{center}
{\Large\bf  Masses and Mixings from Neutrino Beams pointing to Neutrino Telescopes$^*$\footnote{\hspace*{-1.6mm}$^*$Work supported by 
        "Sonderforschungsbereich 375 f\"ur Astro-Teilchenphysik" der 
        Deutschen Forschungsgemeinschaft.}}
\end{center}}
\renewcommand{\thefootnote}{\alph{footnote}}

\vspace*{.8cm}
{\begin{center} {\large{\sc
                K.~Dick\footnote[1]{\makebox[1.cm]{Email:}
                Karin.Dick@physik.tu-muenchen.de}\footnotemark[2],~
                M.~Freund\footnote[3]{\makebox[1.cm]{Email:}
                Martin.Freund@physik.tu-muenchen.de},~  
                P.~Huber\footnote[4]{\makebox[1.cm]{Email:}
                Patrick.Huber@physik.tu--muenchen.de}~and~
                M.~Lindner\footnote[5]{\makebox[1.cm]{Email:}
                Manfred.Lindner@physik.tu--muenchen.de}~
                }}
\end{center}}
\vspace*{0cm}
{\it 
\begin{center}  
        
        \footnotemark[2]${}^,$\footnotemark[3]${}^,$\footnotemark[4]${}^,$\footnotemark[5]%
                Theoretische Physik, Physik Department, 
                Technische Universit\"at M\"unchen,\\
                James--Franck--Strasse, D--85748 Garching, Germany

        \footnotemark[1]%
                Max-Planck-Institut f\"ur Physik, Postfach 401212, D--80805 M\"unchen, Germany 

\end{center} }

\vspace*{1.5cm}

{\Large \bf 
\begin{center} Abstract \end{center} }
We discuss the potential to determine leading oscillation 
parameters, the value and the sign of $\dm{31}$, as well as the magnitude 
of $\sin^2 2\theta_{13}$ using a conventional wide band neutrino beam pointing
to water or ice Cherenkov neutrino detectors known as ``Neutrino 
Telescopes''. We find that precision measurements of $\dm{31}$ and 
$\theta_{23}$ are possible and that, even though it is not possible to 
discriminate  between charges in the detector, there is a remarkably good sensitivity to
the mixing angle $\theta_{13}$ and the sign of $\dm{31}$.

\vspace*{.5cm}

\end{titlepage}

\newpage

\renewcommand{\thefootnote}{\arabic{footnote}}
\setcounter{footnote}{0}


\section{Introduction \label{sec:SEC-intro}}

Recent studies of precision measurements of neutrino mixing parameters
and neutrino masses focused on high intensity beams at
neutrino factories, these offer unique advantages compared to 
conventional wide band neutrino beams. The 
development of a neutrino factory is however very cost intensive and still in its
initial stage. For this reason it was recently emphasized that further studies 
of conventional wide band neutrino beams should be pursued \cite{NUFACT00}. 
One of the major advantages of a neutrino factory 
oscillation experiment would be the possibility to measure the appearance 
oscillation channel $\reub$ (wrong sign muons) in very 
long baseline experiments. Using appearance rates enables the 
determination of the small mixing angle $\theta_{13}$, $|\dm{31}|$ as well as 
the sign of $\dm{31}$, which discriminates between the two possible mass 
ordering schemes \cite{BARGER,FLPR,GAVELA}.
An analysis of the appearance rates requires that the detector can discriminate
the charges of $\mu^{+}$  and $\mu^{-}$ at a sufficient level 
in order to separate wrong sign muons
from the large amount of right sign muons produced in the disappearance channel $\ruu$. 
Therefore, large magnetized iron detectors usually are the detection 
system of choice. In \cite{FHL} it was  however shown that the ability to 
determine $\theta_{13}$ and the sign of $\dm{31}$ does not necessarily depend 
on the capability of charge identification. The information on $\theta_{13}$ 
and the sign of $\dm{31}$ is also contained in the disappearance channel $\ruu$ and 
can thus be extracted from the $\nu_\mu$ rates without charge identification.
If $\theta_{13}$ is not too small a measurement is possible in this channel
with a precision which is comparable to the appearance channel. Furthermore,
such a measurement would even be possible with conventional wide band 
neutrino beams, if a sufficiently high neutrino event rate is achieved, 
which is big enough to limit the statistical error, 
and if the systematic errors on the beam flux are under control. 
We show in this paper that conventional neutrino beams 
(consisting only of $\nu_{\mu}$) 
in combination with large water or ice Cherenkov 
detectors (Neutrino Telescopes without charge identification) 
can be used and give remarkable results.
In particular we use as a 
prototype scenario a CNGS-type beam \cite{CNGS} and an AMANDA-like detector \cite{AMANDA}, 
for which the neutrino event rates are comparable to those of proposed neutrino factory 
experiments. We discuss the problems which arise in measurements of neutrinos in the
energy threshold region of Neutrino Telescopes and suggest using beam pulse
timing information and neutrino direction information to reduce the
background from atmospheric muons. Finally, we perform a numerical analysis
of the physics potential of this type of experiment and show that
precision measurements of the leadings oscillation parameters 
as well as the determination of $\sin^2 2\theta_{13}$, the test of the MSW-effect \cite{MSW}
and the determination of the sign of $\dm{}$ are possible.  


\section{Three Neutrino Oscillations in Matter \label{sec:SEC-phy}}

The basic mechanism which allows the extraction of the sign of $\dm{31}$
comes from coherent forward scattering of electron neutrinos 
in matter (MSW-effect) which leads to effective masses and mixings 
different from vacuum. 
In the approximation where the solar $\dm{}$ is ignored compared to the
atmospheric $\dm{}$, i.e. $\dm{21}=0$ they are given by
$\dm{31,m} = \dm{} C_\pm$, $\dm{32,m} = \dm{} [(C_\pm+1)+ A]/2$,
$\dm{21,m} = \dm{} [(C_\pm-1)- A]/2$ and
$\sin^2 2\theta_{13,m}  = \sin^2 2\theta_{13}/C_{\pm}^2$ where
$C_\pm  = [(A/\dm{} - \cos 2\theta)^2 + \sin^2 2\theta]^{-1/2}$
and $A = 2E V = \pm 2\sqrt{2}G_FY\rho E/m_n$. 
$\theta_{23}$ is not changed in the approximation $\dm{21}=0$
(we follow in this work the notation of \cite{FHL}).
The size of the matter corrections for a given neutrino species depends 
on the sign of $\Delta m_{13}^2$ which enters into $C_\pm$. The resulting 
modification of the muon disappearance probability in matter therefore depends
on the sign of $\Delta m_{13}^2$: 
\begin{eqnarray}
P(\nu_\mu\leftrightarrow\nu_\mu) = 
1 &-& \sin^2 2\theta_{23} \sin^2\theta_{13,m} \sin^2(\Delta_{31,m}) \nonumber  \\
  &-& \sin^4 \theta_{23} \sin^2 2\theta_{13,m} \sin^2(\Delta_{31,m}) \label{eq:mswprob} \\
  &-& \sin^2 2\theta_{23} \cos^2\theta_{13,m} \sin^2(\Delta_{31,m})~. \nonumber 
\end{eqnarray}
For antineutrinos, the matter induced modifications of the
disappearance probability correspond to opposite sign of $\Delta m_{31}^2$ for 
neutrinos. This can be used to amplify the signature of matter effects if $\nu_\mu$ 
and $\bar\nu_\mu$ beams can be compared. The biggest effect from the sign of 
$\Delta m_{13}^2$ will be seen at the MSW resonance energy (in the Earth $\approx$ 10-15~GeV)  
and the sensitivity to  matter effects will thus be best when the maximum of the 
neutrino spectrum is suitably adjusted and the detector is sensitive in this energy 
region. For a more detailed treatment see \cite{FHL}. 

The differential muon event rates in the detector are:
\begin{equation}
\frac{dn_{\mu}}{dE} =\underbrace{10^9~N_A~\NKT~\epsilon_\mu}_{normalization}~ 
\underbrace{\left(\frac{L_{\mathrm{near}}}{L_{\mathrm{far}}}\right)^2 f_{\mathrm{near}}(E)~\sigma(E)~}_{flux}~\underbrace{P_{\ruu}(E)}_{oscillation}~.
\label{eq:differential}
\end{equation}
Here $10^9N_A$ is the number of nucleons per kiloton in the detector,
$\NKT$ is the detector size in kilotons, $\epsilon_\mu$ is the detection 
efficiency and $\sigma$ is the charged current neutrino cross section 
per nucleon. The neutrino beam flux $f_\mathrm{near}$, which we assume to be monitored 
at the near detector, must be corrected by the geometrical suppression 
factor 
$(L_{\mathrm{near}}/L_{\mathrm{far}})^2$, where  $L_{\mathrm{near}}$ and 
$L_{\mathrm{far}}$ are the distances to the near and far detectors.
Total rates are then obtained by integrating these differential 
rates from the threshold to the maximum neutrino energy.


\section{Experimental Issues}
\label{sec:SEC-threshold}

Next we turn to the physics potential of large water or ice Cherenkov detectors 
as components of very long baseline neutrino oscillation experiments. Several 
experiments of this type which consist of photomultipliers which are attached 
to vertical strings are under construction or already taking data 
(AMANDA \cite{AMANDA} / ICECUBE \cite{ICECUBE}, ANTARES \cite{ANTARES}, BAIKAL 
\cite{BAIKAL} and NESTOR \cite{NESTOR}). 
For the proposed ICECUBE array, the vertical spacing is about 15 m,
and the 80 strings of 1~km length are 100~m away from each other.
These Cherenkov detectors allow detector masses up to $10^5$~kt which is a 
factor 100 more than the most ambitious ideas proposed for a 
megaton Super-Kamiokande-like water Cherenkov detector. These detectors, which 
conventionally are called ``Neutrino Telescopes'', are primarily built in order to 
search for very high energy cosmic and atmospheric neutrinos. In combination with  
neutrino beams, some comments are in order concerning 
measurement of neutrinos very near the threshold energy of the detector 
(20 to 100 GeV). For very long baseline oscillation experiments it is important to 
achieve a detector energy threshold of 10~GeV to 20~GeV and reasonable energy 
resolution of about 10~GeV. This low neutrino energy is needed
to take advantage of the MSW-effects which enhances sub-leading effects from 
$\theta_{13}$ and from the sign of $\dm{31}$. The usually quoted energy thresholds 
for the detectors mentioned above are a result of the geometry 
of the photo multiplier (PM) array, the optical properties of ice or sea 
water and the specific energy loss of muons per traveled distance
($\sim 0.2~\mathrm{GeV}/ \mathrm{m}$). Only upward moving muons are
unambiguously assigned to neutrino interactions and the ability to
reconstruct the particle track with  sufficient precision is crucial in order
to reject the huge background from downward going muons. Typically, hits in 7 or more PMs  
are requested  for proper track reconstruction. For AMANDA, the most simple majority trigger 
requests 8 to 16 hit PMs, and this sets the energy threshold to 30-50 GeV. By triggering on
certain hit patterns like hits in 5 neighboring (or nearly neighboring) PMs with some of them 
having high amplitude, one can select muons close to a string, with lower energy threshold
(and of course paying with a reduced effective volume). Still, the
background of downward muons dwarfs the signal, since the tracking error obtained
from 5 hits is rather large.  The situation is however different, when a neutrino beam 
is used as a source: The background can be reduced by five to six orders of magnitude, 
if beam pulse timing information\footnote{The duty cycle of wide band neutrino
beams is about $1:10^6$, that of neutrino factories may reach
$1:10^3$ \cite{GEER98}.
The absolute timing accuracy of events is about 100~ns (GPS accuracy).} 
and beam direction information\footnote{In the case discussed here, the neutrino beam 
will meet the detector under a small angle to the direction of the strings. 
} is used. Independently a lower bound on the threshold is imposed by the 
requirement to distinguish hadronic showers which are produced by neutral current 
reactions of all neutrino flavors. The length of
such a shower grows as $\log E$; the track length of a muon grows as $E$;
therefore a cut on the minimum track-length is used which corresponds to an
energy threshold of 10~GeV to 20~GeV, depending on the optical properties of
the detector medium, and yields an energy resolution in the same range.

These arguments suggest strongly that Neutrino Telescopes can be used as
large mass detectors for very long baseline neutrino experiments, but this
idea requires certainly further investigation. For our work we assume a prototype detector
with an effective mass of 5000~kt. This seems to be a reasonable estimate for the effective
volume of AMANDA and ANTARES with respect to muons in the range of 
10-20 GeV.  The corresponding ICECUBE volume might come close to 10 to
100~Mt. The assumed energy threshold is 15~GeV and the energy resolution is 10~GeV \cite{SPIERING}.
To illustrate the enormous potential of such a detector, we use as source a
conventional wide band neutrino beam like NuMI or CNGS. The CNGS beam consists 
mainly of $\nu_{\mu}$ with an admixture of about 2\% $\bar \nu_{\mu}$, 
0.8\% $\nu_{e}$ and 0.05\% $\bar\nu_{e}$ and the number of additional muons expected 
from oscillated $\nu_e$'s in the beam is small and negligible. The mean energy of the 
beam lies in the range of 20 GeV to 30 GeV, with a long tail towards higher energies. In principle 
it is also possible to get a $\bar\nu_{\mu}$-beam by reversing the current 
in the lens system; this would result however in an smaller flux of about 
75\% and larger admixtures of other neutrinos.


\section{Results
\label{sec:SEC-results}} 

For our analysis we assume a 5000~kt detector with energy threshold 15~GeV, 
energy resolution 10~GeV (corresponding to four bins) and $\epsilon_\mu = 100\%$ 
detection efficiency. As neutrino source we used the CNGS flux spectrum for 
$4.5\cdot 10^{19}$ protons on target \cite{CNGS}. We numerically calculate the 
charged current rate spectrum at the detector according to eq.~\ref{eq:differential}. 
To compute the transition probability we integrate eq.~\ref{eq:mswprob} over the 
full Stacey matter density profile of the Earth \cite{STACEY}. We 
then impose Poisson fluctuations on the rates and re-extract the 
oscillation parameters with a maximum likelihood method
(for details see \cite{FHL}). Systematic errors of the beam
flux and backgrounds are assumed to be small and are not included.
The calculations are performed in the approximation $\dm{21}=0$, i.e. the solar 
mass squared splitting is ignored which is here a very good approximation. 
We investigate two different baselines. First 6500~km (approximately Fermilab--ANTARES), 
to measure $\theta_{13}$ and then 11200~km (CERN--AMANDA), to test the MSW-effect 
and to determine the sign of $\Delta m^2_{31}$. 

\begin{figure}[htb!]
\begin{center}
\epsfig{file=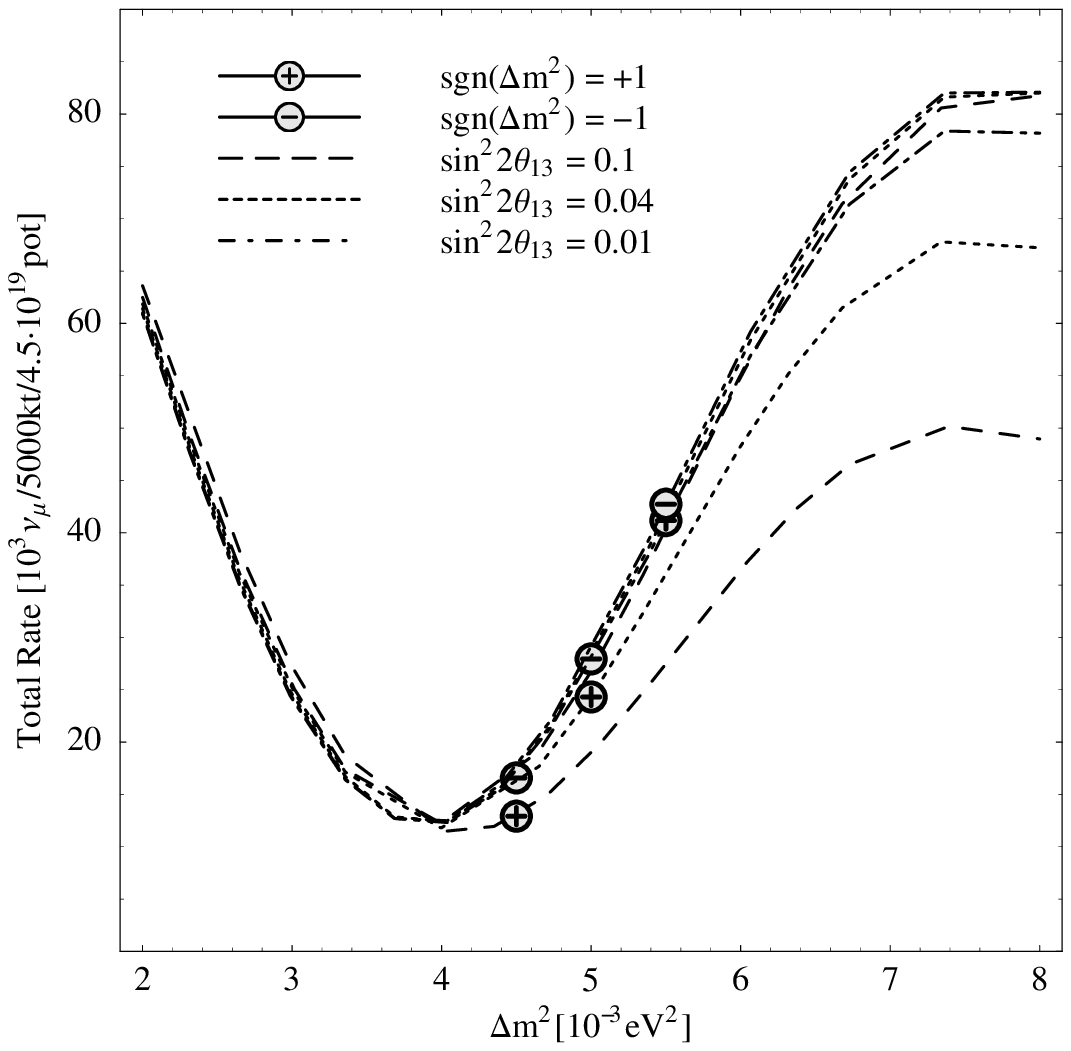,width=9cm}
\end{center}
\mycaption{Total muon rate $\nu_\mu$ as a function 
of $|\dm{31}|$ for a baseline of 6500~km and $\sin^2 2\theta_{23} = 1$.
The different line types stand for different mixing angles $\theta_{13}$
(see labels). $\oplus$ and $\ominus$ indicate positive and negative 
mass squared difference.}
\label{fig:rates}
\end{figure}

In fig.~\ref{fig:rates} the total event rates at a baseline of 6500~km are 
shown as function of the atmospheric mass squared difference $\dm{31}$ for
different values of $\theta_{13}$ and for the two possible signs of $\dm{31}$ 
(labeled $\oplus$ and $\ominus$).  
Apparently, a nonzero $\theta_{13}$ induces a 
significant depletion of the total rates only in the case of a positive sign 
of $\dm{31}$. The reason for this is that the beam consists only of muon neutrinos 
which show in matter MSW-resonant enhancement only for positive $\dm{31}$. In this case, 
studies of the sub-leading oscillation parameter $\theta_{13}$ and the sign of 
$\dm{31}$ give a statistically 
more significant result than with negative sign of $\dm{31}$ (see discussion below).
If indeed the sign of $\dm{31}$ is negative, better significance would be 
achieved with an antineutrino beam, since antineutrinos are resonant in this case. 
It might, however, not be easy to cope with the systematic errors in the beam fluxes
which are substantially different in the charge conjugated channel. 

\begin{figure}[htb!]
\begin{center}
\epsfig{file=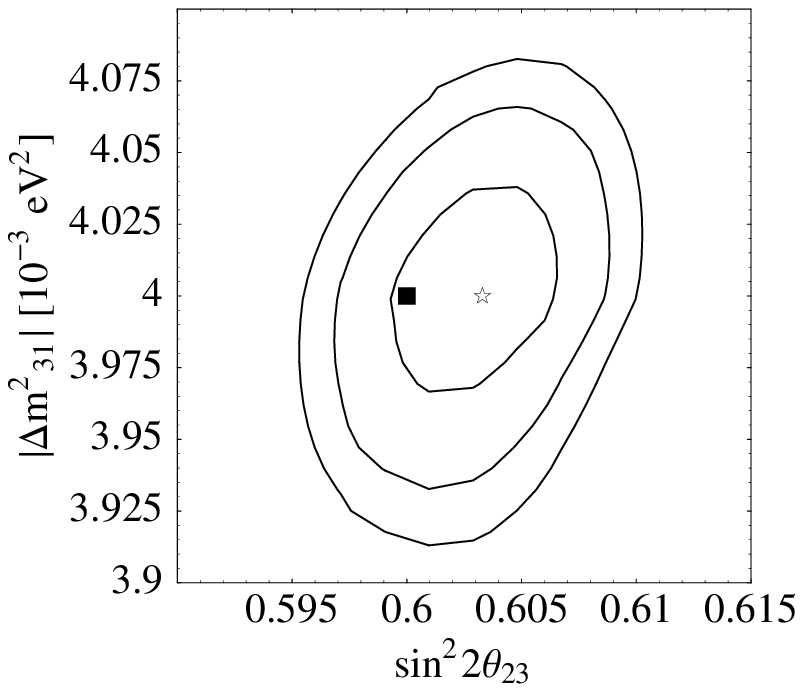,width=7cm}
\end{center}
\mycaption{Fit to the muon neutrino spectrum  (i.e. the unoscillated ($\ruu $) muon 
neutrinos) for $\dm{31}>0$ at a 
baseline of 6500~km. Shown are the $1\sigma$, $2\sigma $ and $3\sigma $ contours. 
The rectangle denotes the parameter pair for which the data are generated 
($\sin^2 2\theta_{23} = 0.6$, $\dm{31} = 4.0\cdot 10^{-3}$) and the star denotes 
the obtained best fit.}
\label{fig:leading}
\end{figure}

\begin{figure}[htb!]
\begin{center}
\epsfig{file=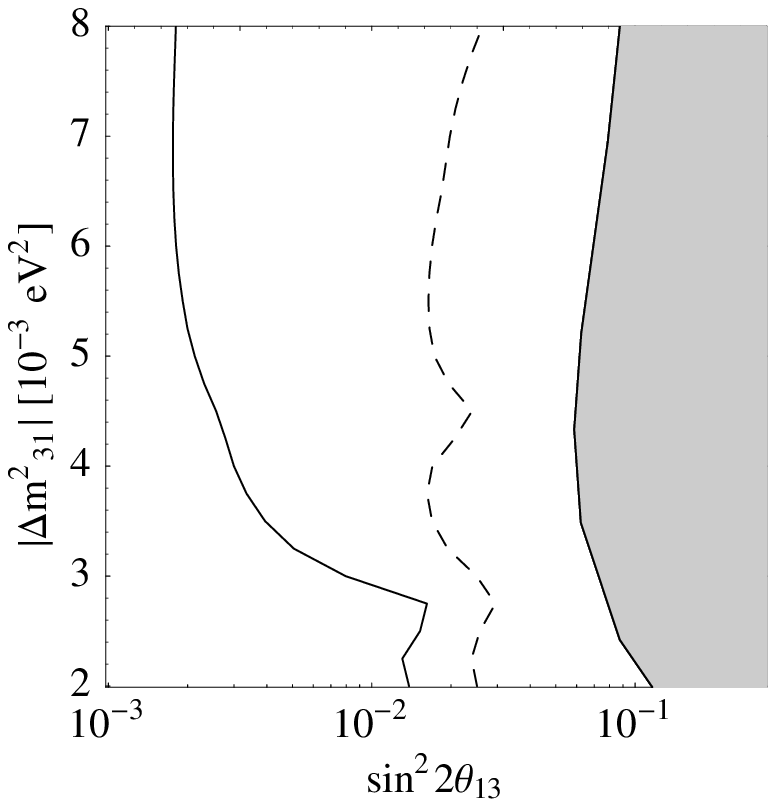,width=7cm}
\end{center}
\mycaption{The 90\% C.L. exclusion line for
$\sin^2 2\theta_{13} \equiv 0$ at a baseline of 6500~km and with $\sin^2 2\theta_{23} = 1$. 
The solid line is for the case $\dm{31}>0$ and the dashed line for
$\dm{31}<0$. The shaded area is excluded by the CHOOZ experiment at 
90\% C.L.}
\label{fig:excl}
\end{figure}

\begin{figure}[htb!]
\begin{center}
\epsfig{file=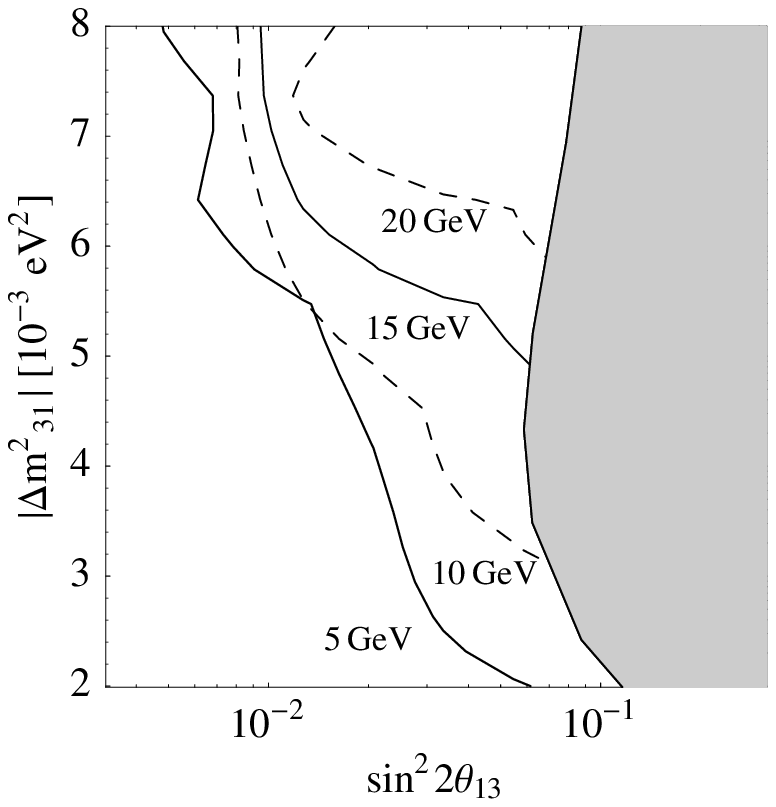,width=7cm}
\end{center}
\mycaption{Sensitivity to the sign of $\dm{31}$ at 90\% C.L. for $\dm{31} > 0$. 
The different lines were obtained with threshold energies of 5~GeV, 
10~GeV, 15~GeV and 20~GeV at a baseline of 11200~km and with 
$\sin^2 2\theta_{23} = 1$. The shaded area is excluded by the CHOOZ experiment at
90\% C.L.}
\label{fig:vzexcl}
\end{figure}

In order to simulate the extraction of the oscillation parameters, we performed 
fits to simulated spectral rates. Instead of performing a global fit 
to all parameters we first fitted the leading oscillation parameters
$\theta_{23}$ and $|\dm{31}|$, which are, to a good approximation, independent
of $\sin^2 2\theta_{13}$. As second step, the sub-leading parameters
$\theta_{13}$ and the sign of $\dm{31}$ are fitted with the leading parameters
fixed to the previously obtained best fit values. The result of the fit
of the leading parameters for a given sample pair of parameters 
($\sin^2 2\theta_{23} = 0.6$, $\dm{31} = 4.0\cdot 10^{-3} \eV^2$) is shown in 
fig.~\ref{fig:leading}. The relative $3\sigma$-errors for the whole parameter 
region which is allowed by the Super-Kamiokande experiment are between 0.5\% and 10\%, 
depending on the value of $\dm{31}$.  
With the result of the fit of the leading parameters, it is possible
to proceed to the second step: the fit of the sub-leading parameters 
$\sin^2 2\theta_{13}$ and the sign of $\dm{31}$. 
Fig.~\ref{fig:excl} shows the parameter region in the $\sin^2 2\theta_{13}$--$\dm{31}$
plane in which the obtained simulated spectral rates are not consistent with 
$\theta_{13} \equiv 0$ at 90\% confidence level. The shaded area is excluded by the 
CHOOZ experiment \cite{CHOOZ}. As explained above, the result differs 
for the two possible signs of $\dm{31}$. For positive $\dm{31}$ (solid line), 
a sensitivity for mixing angles down to $\sin^2 2\theta_{13} \approx 2\cdot10^{-3}$ 
is achieved. For negative $\dm{31}$ (dashed line), the sensitivity is worse and 
does not reach the $\sin^2 2\theta_{13} \approx 10^{-2}$ level.
In both measurements, $\theta_{13}$ and the sign of $\dm{31}$, the use of an antineutrino 
beam would increase the sensitivity to a level comparable to the case of $\dm{31}>0$. 
In fig.~\ref{fig:vzexcl} the $\sin^2 2\theta_{13}$--$|\dm{31}|$ parameter region in which
the determination of the sign of the mass squared difference is possible 
at $90\%$~C.L for a positive sign of $\dm{31}$ is shown. This shows also
the limit of $\sin^2 2\theta_{13}$ where verification of the MSW-effect is possible. 
Since it is very important for this measurement to include neutrinos at the MSW resonance
energy ($\approx$~15~GeV), the value of the threshold of the neutrino detector has a 
substantial influence on the obtained sensitivity. The lines shown in the plot
were obtained with threshold energies of 5~GeV, 10~GeV, 15~GeV and 20~GeV.
A negative $\dm{31}$ in combination with a $\nu_\mu$ beam does not produce MSW-enhanced 
effects in the rates and thus does not allow to determine the sign. If $\dm{31}$ is indeed negative, the 
determination of the sign of $\dm{31}$ and the test of matter effects requires
an anti neutrino beam.


\section{Conclusions \label{sec:concl}}

We studied in this paper the physics potential of large water or ice 
Cherenkov detectors (Neutrino Telescopes) in very long baseline
accelerator neutrino oscillation experiments. In particular we have shown
that conventional wide band neutrino beams pointed to detectors like
AMANDA/ICECUBE, ANTARES or NESTOR give neutrino event rates at the level
of $10^4$ to $10^6$ events per year. This number is comparable to rates 
achieved in presently proposed neutrino factory experiments. 
Neutrino Telescopes are primarily built for the search for ultra high 
energetic cosmic and atmospheric neutrinos. Conventional neutrino beams
provide neutrinos of a few times 10~GeV, which is roughly at the quoted
threshold of Neutrino Telescopes. To cope with this problem, we suggest to use 
beam pulse timing information and neutrino direction information to reduce the 
background produced by cosmic muons in a very effective way. The ultimate threshold 
is then given by limitations which arise from misidentification of hadron showers 
which are produced by all active neutrino flavors. We estimate that an energy 
threshold between 10~GeV and 20~GeV and an energy resolution between 10~GeV and 20~GeV
would finally be achievable. This would open 
the door to a wide spectrum of interesting oscillation physics. 

Our numerical study was performed in the standard three neutrino
scenario under the approximation $\dm{21} = 0$ and taking into account
the full Stacey Earth density model. It demonstrates that precision 
measurements of the leading oscillation parameters 
$\dm{31}$ and $\sin^2 2\theta_{23}$ are possible. 
Using the CNGS beam spectrum as prototype neutrino source, the relative 
$3\sigma$-errors for the whole parameter region which is allowed by the 
Super-Kamiokande experiment are between 0.5\% and 10\%, depending on the value 
of $|\dm{31}|$. We further have demonstrated that, even though a measurement of the 
appearance oscillation channel is not possible (due to the missing capability
to identify charges), there is good sensitivity to the sub-leading oscillation 
parameters $\sin^2 2\theta_{13}$ and the sign of the mass squared difference.
Taking into account only statistical errors, we calculated that 
a measurement of $\theta_{13}$ would be possible down to 
$\sin^2 2\theta_{13}$ values of $2\cdot 10^{-3}$ 
. In case of a negative sign of $\dm{31}$ the sensitivity would 
be worse, but this could be overcome by using an antineutrino beam. The values
given above were all obtained at a baseline of 6500~km. With larger baselines,
a test of the MSW-effect would be possible. In particular, at 11200~km the 
the sign of $\dm{31}$ (which can be revealed only through 
matter effects) can be determined down to $\sin^2 2\theta_{13}$ values of 
approximately $10^{-2}$, depending on the precise value of $|\dm{31}|$.

An important aspect of the experimental scenario studied in this work is that 
a test of the MSW effect and the determination of $\sin^2 2\theta_{13}$ could 
be done with existing technologies. This type of high rate neutrino experiment 
might thus be an interesting alternative to neutrino factory experiments. 
Further studies of systematic errors in the beam flux and backgrounds are however 
necessary and we recommend that detailed simulations of the detector response 
should be performed. A study of the potential of $\nu$-factory beams 
pointing to a km$^3$ Cherenkov detector will be published soon \cite{DFHL2}.


\vspace*{7mm}

{\bf Acknowledgments:}
We thank M.~Leuthold and C.~Spiering for supplying us with
crucial information on detector issues. Furthermore, we want to thank 
L.~Oberauer for helpful discussions. This work was supported by the 
``Sonderforschungsbereich~375 f\"ur Astro-Teilchenphysik'' der Deutschen 
Forschungsgemeinschaft.


\newpage

\bibliographystyle{phaip}

\end{document}